\documentclass[global,twocolumn]{svjour}
\usepackage{graphicx}
\journalname{Applied Physics B}
\begin{document}
\title{Beaming effect from increased-index photonic crystal waveguides}
\author{Steven K. Morrison and Yuri S. Kivshar
}                     
%
%
\institute{Nonlinear Physics Centre and Centre for Ultra-high
bandwidth Devices for Optical Systems (CUDOS),\\ Research School
of Physical Sciences and Engineering,\\ Australian National
University, Canberra, ACT 0200, Australia\\
(Fax: +61-26125-8277, Email:skm124@rsphysse.anu.edu.au)}
\date{Received: date / Revised version: date}
%
\maketitle
\begin{abstract}
We study the beaming effect of light for the case of
increased-index photonic crystal (PhC) waveguides, formed through
the omission of low-dielectric media in the waveguide region. We
employ the finite-di\-ffe\-ren\-ce time-domain numerical method
for characterizing the beaming effect and determining the
mechanisms of loss and the overall efficiency of the directional
emission. We find that, while this type of PhC waveguides is
capable of producing a highly collimated emission as was
demonstrated experimentally, the inherent characteristics of the
structure result in a restrictively low efficiency in the coupling
of light into the collimated beam of light.\\

\hspace{-0.5cm}\textbf{PACS}42.70.Qs; 78.20.Bh \vspace{-0.2cm}
\end{abstract}
\section{Introduction}
\label{intro}
Photonic crystals (PhC) are expected to play an important role in
the development of small integrated optical circuits, combining
the diverse functionality of optical devices and intra-connections that
confine light on a sub-wavelength scale.  Yet, associated with
this sub-wave\-length confinement of light is the complexity of
interfacing the small light waveguides and cavities of photonic
crystals with conventional optical systems such as fibers,
waveguides, and freely propagating light beams. Indeed, coupling
light directly out of PhC waveguides into free-space in an usable
manner is particularly challenging due to the strong diffraction
of light by the sub-wavelength dimensions of the waveguide
exit~\cite{Bethe1944}. However, the \emph{beaming effect of
light}, which has been studied in metallic-thin-film
systems~\cite{Ebbesen_N_98,Lezec_S_02} and shown to exist in
PhCs~\cite{Moreno_PRB_04,Moreno_PNFA_04,Kramper_PRL_04}, has been
suggested as a possible approach to overcome these limitations,
allowing highly directed emissions from PhC waveguides that exist
below the diffraction limit. To overcome the diffraction limit and
achieve a directed emission, the beaming effect utilizes leaky
surface modes and coherent interference to redistribute the power
of the transmitted light into a narrow beam directly in front of
the waveguide exit. Coupling from the leaky or radiative surface
modes is achieved through a periodic corrugation in the exit
surface of the PhC structure, with the geometric and material
properties of this corrugation establishing a spatial phase and
amplitude distribution within the radiated field that, under
appropriate conditions, leads to a highly directed emission.

However, the full utility and efficiency of the beaming effect is
yet to be determined. Our recent study has illustrated the
potential to substantially enhance and control the beaming effect
by engineering the surface and near-surface structure of a
particular PhC~\cite{SKM_APL_05}. In this case, the enhanced
beaming structure makes use of a PhC created using a square
lattice of high dielectric rods in air, where a row of rods is
removed to form a waveguide. Unfortunately, this type of PhC does
not exhibit a complete photonic band gap (PBG) for both
polarisations of light and the surface structure required for the
beaming effect is not easy to fabricate. For these reasons, in
this paper we examine the effectiveness of the beaming effect in
the more readily manufactured PhC structure---a triangular lattice
of holes in a high dielectric material---in which the beaming
effect has been experimentally demonstrated~\cite{Kramper_PRL_04}.
A waveguide can be formed within this PhC through the omission of
a row of holes to create an increased-index guiding region that
does exhibit a PBG for both polarisations of light. In this paper,
we analyze, by the finite-difference time-domain numerical method,
the conditions required to achieve optimal beaming from the
increased-index waveguide, highlighting the sources of losses and
inefficiencies. From our analysis, we illustrate how this structure,
while producing a highly directed emission, does not achieve an
efficient coupling to the transmitted light, thus limiting its
technological application.

\section{Characterisation of the directional emission}
\label{sec:2}

\begin{figure}[t]
\begin{center}
\includegraphics[width=9cm,keepaspectratio=true]{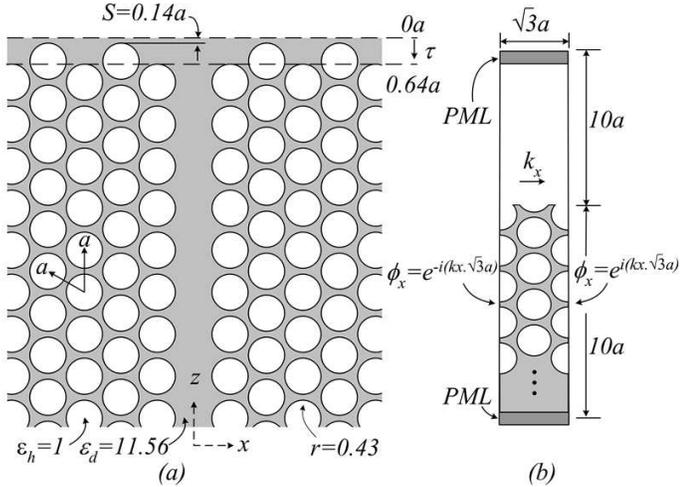}
\caption{Photonic crystal model based on Ref.
\cite{Kramper_PRL_04,Schilling_JOO_01}: (a) detailing the
waveguide and surface structures; and (b) depicting the supercell
construction for calculation of the surface modes' dispersion
relationships.} \label{Fig1}
\end{center}
\end{figure}

We consider the PhC model based on the experimentally demonstrated
structures described, for example, in \mbox{Refs.
\cite{Kramper_PRL_04,Schilling_JOO_01}}. It consists of a
two-dimensional triangular lattice of holes created in a
background of high dielectric material with a dielectric constant
of $\epsilon_{r}=11.56$, representing silicon at a wavelength of
$1.5\mu m$. With a hole-radius to lattice-pitch ratio of
$r/a=0.43$, a photonic band gap is created for TM polarized light
(magnetic field parallel to the hole and travelling in a plane
perpendicular to the holes) within the frequency range of
$\omega=0.28\times 2\pi c/a$ to $0.45\times 2\pi c/a$. We
orientate the triangular lattice such that the $\Gamma-K$
irreducible Brillouin vector is directed along the $z$-axis, and
form a waveguide in this direction at $x=0$, through the omission
of a row of holes, as depicted in Fig. 1(a). A Gaussian source is
introduced to the waveguide $20a$ from the waveguide exit to
ensure only true waveguide modes are coupled from the guide. The
dispersion relationships for the even-symmetry modes of this
waveguide are presented in Fig. 2. The terminating surface of the
PhC is formed along a plane parallel to the $\Gamma-M$ direction,
perpendicular to the waveguide. This terminating plane, introduced
anywhere within an infinite PhC, leaves partial holes along the
surface resulting in a natural surface corrugation, providing wave
vector matching to achieve coupling between surface and radiative
modes.
\begin{figure}[h]
\begin{center}
\includegraphics[width=8cm,keepaspectratio=true]{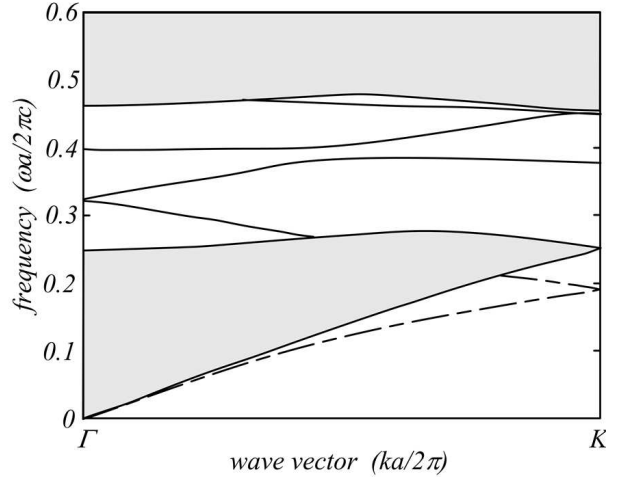}
\caption{Project dispersion relationship of the even-symmetry
waveguide modes. A mini-stop band exists between the second and
third waveguide bands.} \label{Fig2}
\end{center}
\end{figure}
\begin{figure}[b]
\begin{center}
\includegraphics[width=8.7cm,keepaspectratio=true]{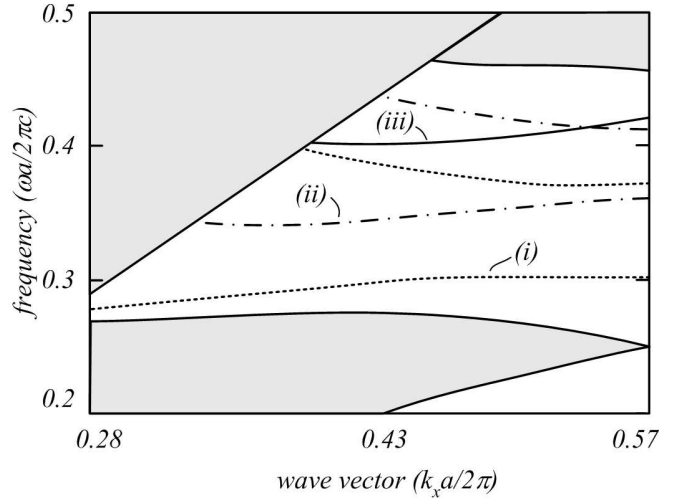}
\caption{Surface mode dispersion relationship within the
dielectric in-fill layer for three surface termination of: (i)
$\tau=0$; (ii) $\tau=0.1a$; and (iii) $\tau=0.2a$} \label{Fig3}
\end{center}
\end{figure}

To determine the surface terminations that will support surface
modes, a supercell representation of the surface is created with a
combination of periodic boundary and perfectly matched layer (PML)
boundary conditions, as illustrated in Fig. 1(b). The fields
within the supercell are represented as complex numbers, allowing
the spatial phase to be varied over the surface Brillouin zone as
the eigenmodes of the surface structure are determined from
Fourier analysis.  A small surface termination region exists that
supports surface modes, however, the modes of this region are
weakly localized, causing them to be rapidly radiated. This rapid coupling of the surface modes to the radiated
field is a result of the sharp cusps of high dielectric created by
the partial holes along the surface [see surface of Fig. 1(b)],
resulting in an optical rough surface that limits the spatial
distribution of the diffractively focussing components of light,
thus limiting the formation of the directed emission. To obviate
this limited distribution of the surface modes, we apply a
dielectric layer to the truncated surface that in-fills the
partial holes, as depicted in Fig. 1(a). This in-fill layer has
two effects: it reduces the roughness of the surface allowing the
surface modes to travel further along the surface, and it changes
the surface corrugation period within the range of surface
terminations that support surface modes. Again, using the
supercell method we calculate the surfaces that support
surface modes within the dielectric in-fill region, defined by
$\tau$ in Fig. 1(a). Figure 3 illustrates the dispersion
relations for three surface terminations ($\tau=0, 0.1a,
0.3a$) within this region.
\begin{figure*}[t]
\begin{center}
\includegraphics[width=17.5cm,keepaspectratio=true]{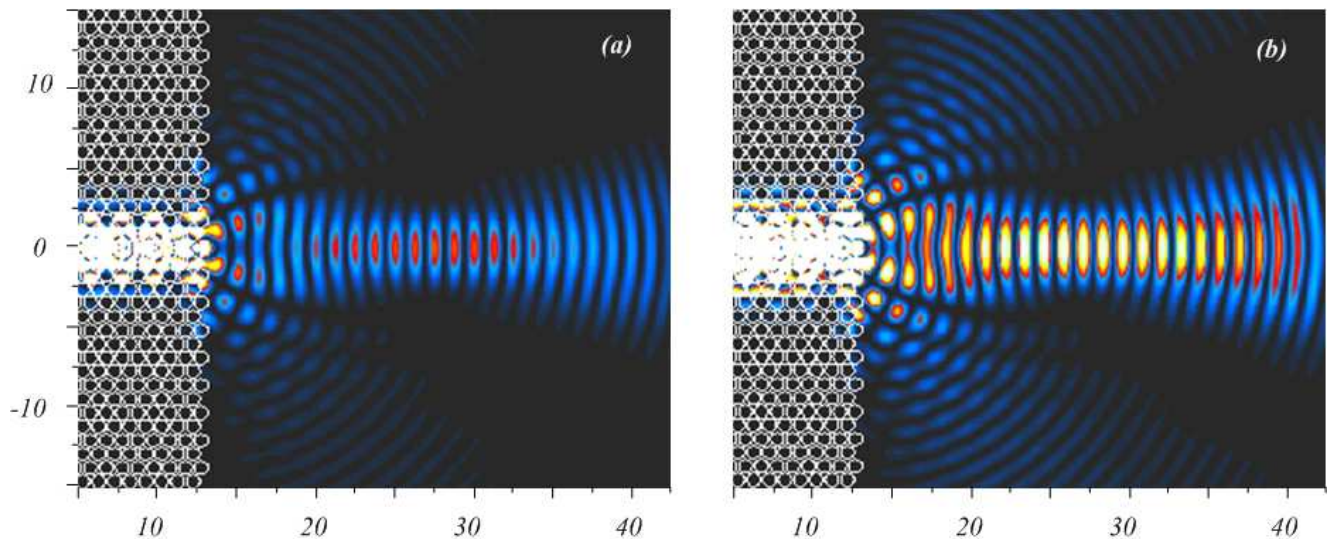}
\caption{Spatial distribution of the intensity from the
increased-index waveguide for: (a) a surface termination of
$\tau=0.2a$ and frequency $\omega = 0.409 \times 2\pi c/ a$; and
(b) a surface termination of $\tau=0.2a$ and frequency $\omega
=0.409 \times 2\pi c/ a$, with a Fabry-Perot resonance induced by
terminating the input end of the waveguide with the bulk of the
photonic crystal and the other end by the partially reflective
waveguide--vacuum interface.} \label{Fig4}
\end{center}
\end{figure*}

A spectral analysis of a series of terminations within the surface
mode forming region reveals that optimal beaming occurs for a
surface termination of $\tau=0.2a$ at frequencies corresponding to
the band-edges of the waveguide's mini-stop
band~\cite{Agio_PRE_01} [see Fig.2], where the waveguide mode's
group velocity tends to zero, increasing the coupling time to
surface modes and consequently increasing the power within the
diffractively focussing components of light.To characterize this
optimal directed transmission we calculate the fraction of power,
normalized to the total power, within the central lobe of the
directed emission in the simulation domain and between the first
nulls of the beam, $P_{d}$, and the width of the beam between the
first null, $w_{d}$, at a distance of $20a$ in front of the
waveguide exit. A likewise normalized measurement is also taken of
the combined power within the surface modes and transmitted side
lobes, $P_{s}$.  In addition to these measurements, we calculate
the return loss, $R$, expressed at a fraction of the total power,
experienced at the waveguide exit. In this manner, we characterize
the directed emission for a surface termination of $\tau=0.2a$ and
source frequency $\omega=0.409 \times 2\pi c/a$ to have a directed
power of $P_{d}=0.054$, a return loss of $R=0.93$, with the
remaining power of $P_{s}=0.016$, delivered into the surface modes
and side lobes. In this configuration, the directed emission has a
beam width of $w_{d}=6.2a$, as illustrated in the color contour
plot of the spatial light intensity of Fig. 4(a), where the color
scaling highlights the beaming. Notably, 77\% of the transmitted
light emitted from the waveguide is delivered into the highly
collimated emission; unfortunately though, this accounts for only
5.4\% of the total light. The very poor transmission of light from
the waveguide is a result of large Fresnel reflections occurring
at the waveguide-vacuum interface as the light attempts to move
from a high refractive index to a low refractive-index media. The
transmission is also further limited by an impedance mismatch
between the waveguide mode and the surface and radiated fields.

Noting that optimal beaming occurred for a waveguide mode with
vanishing group velocity, and that the waveguide exit forms a
partially transmitting mirror, we take the findings to their
natural extension: a Fabry-Perot resonant cavity. To complete the
basic elements of the cavity, we terminate the internal end of the
waveguide with the bulk of the photonic crystal, and tune the
cavity length to a Fabry-Perot resonance. Using the previously
described characterization method, we find the direct emission is
doubled, the power within the surface modes and side lobes equally
doubled, and the beam width maintained. The reflection coefficient
at the waveguide exit is expected to be similar to that of the initial
structure, as there is minimal change to the waveguide mode
structure. The spatial distribution of the field intensity for the
Fabry-Perot cavity beaming is illustrated in Fig. 4(b), with a
plot of the power cross-section highlighting the beaming profile
of the two structures given in Fig. 5.
\begin{figure}[t]
\begin{center}
\includegraphics[width=8.5cm,keepaspectratio=true]{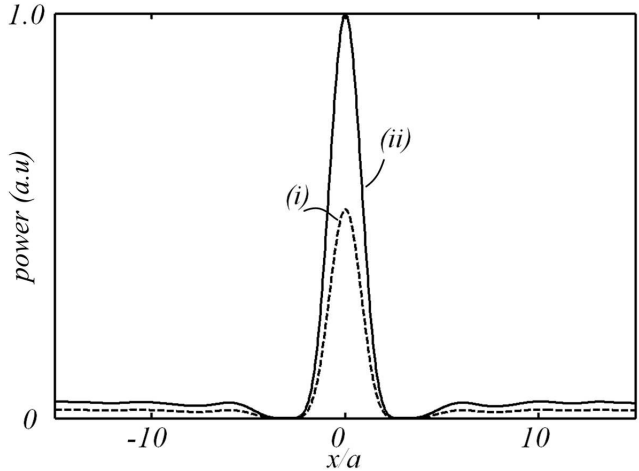}
\caption{Power density incident upon a cross-section $20a$ in
front of the waveguide exit for: (i) optimal beaming conditions
[see text]; and (ii) optimal beaming conditions from a Fabry-Perot
resonant cavity. } \label{Fig5}
\end{center}
\end{figure}

\section{Discussions}
\label{sec:3}

We believe that the primary difficulty in achieving highly
efficient coupling of light into the directional emission is the
large Fresnel reflections that occur from the step change in the
refractive index at the waveguide-vacuum interface. As this step
change is fundamental to the waveguide structure, little can be
done to reduce this source of inefficiency.  Minor reduction of
the reflections may be possible for different mode symmetries at
different bands of the dispersion relationship or with the
introduction of defects within the waveguide and near the exit, as
has been demonstrated in the experimental
setup~\cite{Kramper_PRL_04}. However, in introducing these changes
the other conditions from optimal beaming---low group velocity and
source frequency matching to the surface corrugation
period---would change, requiring further engineering of the
structure. Indeed, the large reflections at the waveguide exit are
not the only issue of a possible concern. The narrow bandwidth for
optimal beaming is also a restrictive attribute; limiting the
information-carrying capacity of light transmitted from the
waveguide. Indeed, in its present form, the low bandwidth capacity
of the beaming effect would prevent it from being used in control
or communication applications, with this restriction being general
to the beaming effect's mechanism of operations, and not just the
PhC and waveguide structure considered here.

However, as demonstrated by the Fabry-Perot cavity beaming, the
beaming effect can be used to collimate light from cavity defects,
opening the possibility for controlling and shaping the emissions
from near-surface light sources such as light emitting diodes and
lasers.  Indeed, the mechanism for determining optimal
beaming---low group velocity within the cavity to improve the
coupling time to surface modes---reinforces this application.

\section{Conclusions}
\label{sec:3}

We have analyzed the conditions for optimal beaming of light from
an increased-index waveguide formed within a PhC structure created by a
triangular lattice of holes in a high dielectric material. From
this analysis, we have demonstrated the inefficiencies that result
from large Fresnel reflections that occur at the waveguide-vacuum
interface, and how this loss mechanism will be fundamental to all
such high-index waveguide structures used to produce the beaming
effect, thus limiting the technological applications of this type
of PhC structure.

\section*{Acknowledgements}
\label{sec:3}

We acknowledge the support of the Australian Research Council through the Centre of Excellence Program and
useful discussions with Sergei Mingaleev and Costas Soukoulis.

\bibliographystyle{unsrt}
\bibliography{SKM_YSK_MS3182B}

\end{document}